\PassOptionsToPackage{backref,breaklinks=true,colorlinks,citecolor=blue,urlcolor=[rgb]{0.0,0.0,0.8}}{hyperref}
\documentclass[iop,apj,appendixfloats,twocolappendix]{emulateapj}
\usepackage{adjustbox}
\usepackage{multirow}
\usepackage{colortbl}
\usepackage[fleqn]{amsmath}
\usepackage{enumerate}
\usepackage{float}
\usepackage{placeins}
\usepackage{hyperref}
\usepackage[all]{hypcap}
\defcitealias{kalap2014}{KHK}
\shorttitle{Understanding the Pulsar Gamma-Ray Emission}
\shortauthors{Kalapotharakos et al.} 

\begin{document}

\title{\emph{Fermi} Gamma-Ray Pulsars:\\
Understanding the High-Energy Emission\\ from Dissipative
Magnetospheres}

\author{Constantinos Kalapotharakos}
\affil{Universities Space Research Association (USRA) Columbia, MD
21046, USA} \affil{Astrophysics Science Division, NASA/Goddard Space
Flight Center, Greenbelt, MD 20771, USA} \affil{University of
Maryland, College Park (UMDCP/CRESST), College Park, MD 20742, USA}
\email{constantinos.kalapotharakos@nasa.gov}
\email{ckalapotharakos@gmail.com}

\author{Alice K. Harding}
\affiliation{Astrophysics Science Division, NASA/Goddard Space
Flight Center, Greenbelt, MD 20771, USA}

\author{Demosthenes Kazanas}
\affiliation{Astrophysics Science Division, NASA/Goddard Space
Flight Center, Greenbelt, MD 20771, USA}

\author{Gabriele Brambilla}
\affiliation{Universities Space Research Association (USRA)
Columbia, MD 21046, USA} \affiliation{Astrophysics Science Division,
NASA/Goddard Space Flight Center, Greenbelt, MD 20771, USA}
\affiliation{Dipartimento di Fisica, Universit\`{a} degli Studi di
Milano, Via Celoria 16, 20133 Milano, Italy} \affiliation{Istituto
Nazionale di Fisica Nucleare, Sezione di Milano, Via Celoria 16,
20133 Milano, Italy}

\slugcomment{Accepted 5/2/2017}



\begin{abstract}

\noindent Based on the \emph{Fermi} observational data we reveal
meaningful constraints for the dependence of the macroscopic
conductivity $(\sigma)$ of dissipative pulsar magnetosphere models
on the corresponding spin-down rate, $\dot{\mathcal{E}}$. Our models
are refinements of the FIDO (Force-Free Inside, Dissipative Outside)
models whose dissipative regions are restricted on the equatorial
current-sheet outside the light-cylinder. Taking into account the
observed cutoff-energies of all the \emph{Fermi}-pulsars and
assuming that \textbf{a)} the corresponding $\gamma-$ray pulsed
emission is due to curvature radiation at the
radiation-reaction-limit regime and \textbf{b)} this emission is
produced at the equatorial current-sheet near the light-cylinder, we
show that the \emph{Fermi}-data provide clear indications about the
corresponding accelerating electric-field components. A direct
comparison between the \emph{Fermi} cutoff-energies and the model
ones reveals that $\sigma$ increases with $\dot{\mathcal{E}}$ for
high $\dot{\mathcal{E}}$-values while it saturates for low ones.
This comparison indicates also that the corresponding gap-width
increases toward low $\dot{\mathcal{E}}$-values. Assuming the
Goldreich-Julian flux for the emitting particles we calculate the
total $\gamma-$ray luminosity $(L_{\gamma})$. A comparison between
the dependence of the \emph{Fermi} $L_{\gamma}$-values and the model
ones on $\dot{\mathcal{E}}$ indicates an increase of the emitting
particle multiplicity with $\dot{\mathcal{E}}$. Our modeling guided
by the \emph{Fermi}-data alone, enhances our understanding of the
physical mechanisms behind the high energy emission in pulsar
magnetospheres.

\end{abstract}

\keywords{pulsars: general---stars: neutron---Gamma rays: stars}



\section{Introduction}

Pulsars are among the most powerful and robust electromagnetic
machines in the Universe that operate in extreme physical conditions
producing low-frequency electromagnetic (EM) waves ($<3\rm kHz$) and
particle radiation that covers the entire EM spectrum. The machine
(energy) fuel is their huge rotational kinetic-energy $(\sim
10^{45}-10^{52} \rm ergs)$ while their enormous surface
magnetic-field $(B_{\star}\sim 10^{8} \rm{~and~} 10^{13} \rm G)$
mediate the conversion of this energy into the observed particle
radiation.

\emph{Fermi} has played a catalytic role in the current modeling of
the high-energy emission in pulsar magnetospheres. Since its lunch
in 2008 the number of the detected $\gamma-$ray pulsars has
increased by a factor of 30. Thus, now more that 200 $\gamma-$ray
pulsars have been detected (117 of them are compiled in the second
pulsar catalog (2PC); \citealt{2013ApJS..208...17A}). This has
shifted the study of $\gamma-$ray pulsars from discovery to
astronomy by establishing a number of trends and correlations.

Even though the general principles that govern the pulsar
``machine'' have been known for decades the detailed physical
mechanisms that provide a complete interpretation of the
observations remain unknown. The numerical Force-Free (FF) and
magnetohydrodynamical solutions that appeared in the literature over
the past eighteen years for the aligned (2.5D) rotator
\citep{1999ApJ...511..351C,2005PhRvL..94b1101G,
2006MNRAS.368.1055T,2006MNRAS.367...19K,2006MNRAS.368L..30M,
2012MNRAS.423.1416P,2016MNRAS.455.4267C} and for the oblique (3D)
rotators
\citep{2006ApJ...648L..51S,2009A&A...496..495K,2012MNRAS.424..605P,
2013MNRAS.435L...1T} provided the impetus for the exploration of the
field-structure and the properties of more realistic configurations
(compared to the analytic Vacuum-Retarded-Dipole solution (VRD);
\citealt{deutsch1955}).

Although the FF models are probably good indicators of the
magnetic-field-structure, they say nothing about the necessary
accelerating electric-field components $E_{\rm acc}$, which are by
definition zero $(E_{\rm acc}=0)$. \cite{2012ApJ...749....2K} and
\cite{2012ApJ...746...60L} started the exploration of the properties
of dissipative solutions that cover the entire spectrum of solutions
between the VRD and FF ones. In this approach, each adopted
prescription for the current-density incorporates a conductivity
$\sigma$ that regulates the $E_{\rm acc}$. The FF (VRD) solutions
correspond to the $\sigma\rightarrow\infty$ ($\sigma\rightarrow 0$)
regimes.

\cite{2012ApJ...754L...1K} and \citet[hereafter KHK]{kalap2014}
employed these dissipative magnetosphere models to generate model
$\gamma$-ray light-curves due to curvature-radiation (CR). These
studies revealed that the high-$\sigma$ (uniformly distributed)
models place the emission at large distances near the equatorial
current-sheet (ECS) where the demand for the current is high.
Assuming that the radio emission originates near the
stellar-surface, \citetalias{kalap2014} constrained their models
using the observed dependence of the phase-lags between the radio
and $\gamma-$ray emission ($\delta$) on the $\gamma-$ray
peak-separation ($\Delta$). They found that a hybrid form of
conductivity, specifically, infinite conductivity interior to the
light-cylinder (LC) and high but finite conductivity on the outside
provides a significant improvement in fitting the
($\delta-\Delta$)-data. In the so-called FIDO (FF Inside Dissipative
Outside) models, the $\gamma-$ray emission is produced in regions
near the ECS but is modulated by the local physical properties.

In \cite{2015ApJ...804...84B}, we started an exploration of the
spectral properties of the FIDO models. In our study, we used
FF-geometry and approximate $E_{\rm acc}$-values. We tried to find
model-parameters that fit eight bright-pulsars that have published
phase-resolved spectra. The $\sigma$-values that best describe each
of these pulsars showed an increase with the spin-down rate
$\dot{\mathcal{E}}$ and a decrease with the pulsar age.

In this paper, we demonstrate that the information needed to
determine the $E_{\rm acc}$-values (i.e. $\sigma$) is contained on
the \emph{Fermi} cutoff energies $\epsilon_{\rm cut}$, and reveals
also a dependence of  $E_{\rm acc}$ on $\dot{\mathcal{E}}$.
Moreover, we further specify the assumptions of the FIDO models. The
comparison with the \emph{Fermi}-data exposes tight constraints on
the $\sigma$-values uncovering their dependence on
$\dot{\mathcal{E}}$. Finally, this comparison provides clear hints
about the dependence of the corresponding gap-widths and the
multiplicity of the emitting particles on $\dot{\mathcal{E}}$.

\begin{figure}[!tbh]
\vspace{0.0in}
  \begin{center}
    \includegraphics[width=1.0\linewidth]{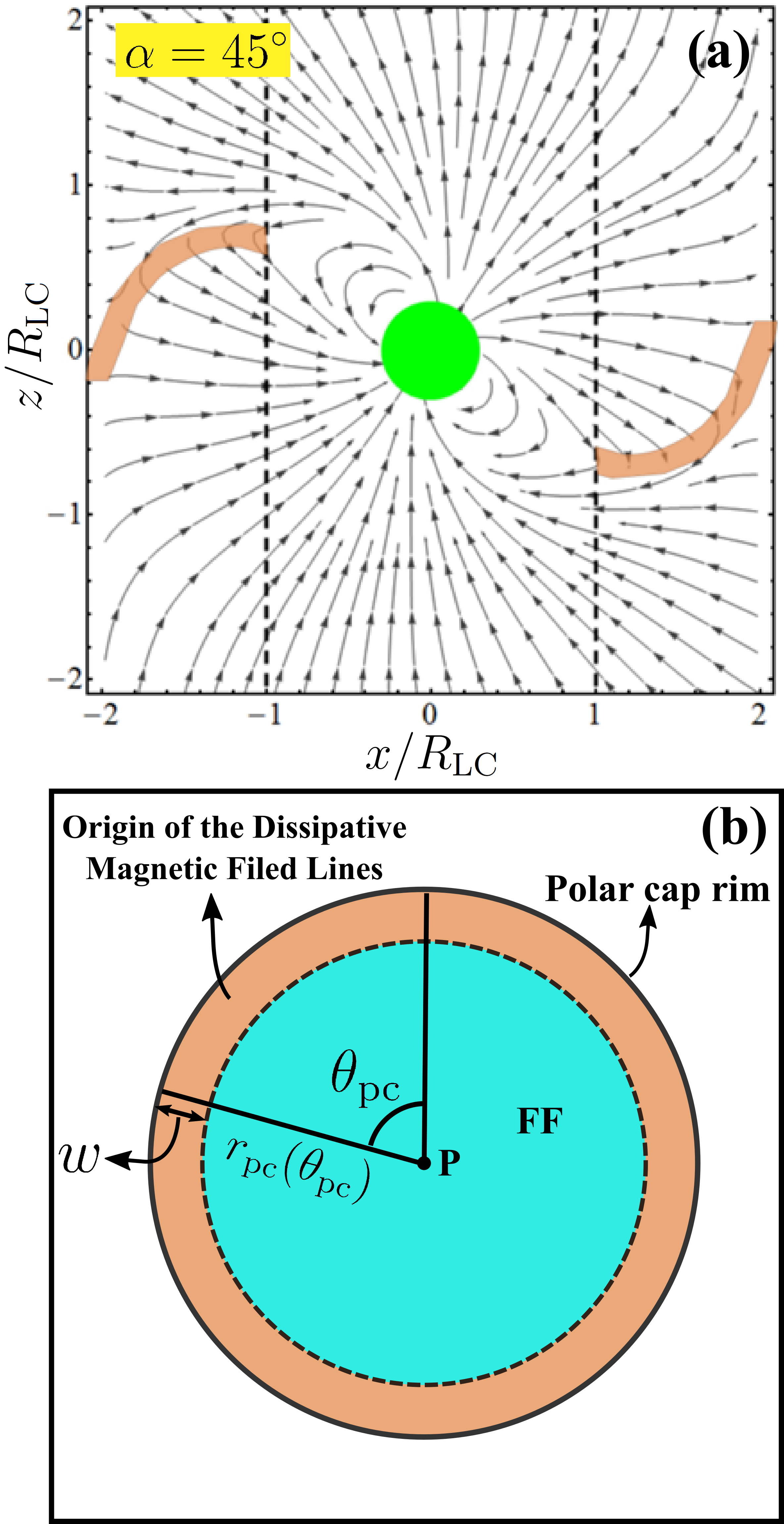}
  \end{center}
  \vspace{0.0in}
  \caption{\textbf{(a)} The dissipative zone (light-orange) is near the ECS beyond the LC.
  \textbf{(b)} The origin of the magnetic-field-lines of the dissipative (light-orange)
  and FF zone (light-blue) on the polar-cap. The gap-width $w$ is a fraction of
  the  polar-cap radius $r_{\rm pc}$.}
  \label{fig01}
  \vspace{-0.1in}
\end{figure}

\section{FIDO model revisited}\label{sec2}

The FIDO model postulates that the magnetospheric plasma
conductivity is finite only outside the LC. For solutions near the
FF ones the adopted approximated expressions used in
\citetalias{kalap2014} and \cite{2015ApJ...804...84B} produce
significant $E_{\rm acc}$-values only near the ECS. These studies
indicated also the necessity of low $\sigma$-values even though the
FF assumption implies only high $\sigma$. Nonetheless, we have found
that the application of small $\sigma$-values everywhere outside the
LC destroys the global FF-field structure (especially for low
inclination-angles $\alpha$) whose geometric properties are
necessary for the successful reproduction of the $\delta-\Delta$
correlation. The only way to keep the field-structure near the FF
one is to apply the low-$\sigma$ in a narrow-zone near the ECS
outside the LC (i.e. near the open-field-boundary). {This actually
implies that the conductivity is small in places the requirement for
the current is high. However this approach requires the detailed
determination of the polar-cap rim at each time-step of the
simulation since the exact 3D locus of the ECS is not a priori
known.} Nonetheless, we have incorporated this into our code which
is now able to apply different $\sigma$-values (in the
current-density prescription shown in eq.~9 of
\citetalias{kalap2014}) along different magnetic-field-lines.

In Fig.~\ref{fig01}a we show schematically the dissipative region
(i.e. finite-$\sigma$). In the light-orange region a finite-$\sigma$
has been applied while all the other regions are FF
($\sigma\rightarrow\infty$). Numerically, the FF condition is
achieved by integrating Maxwell's equations using a high-$\sigma$
($\approx 10\Omega$; where $\Omega$ is the stellar
angular-frequency) and nulling any remaining $E_{\rm acc}$ (only
inside the FF region) at the end of each time-step; this ensures no
parallel electric-component ($E_{\parallel}$) and $E<B$
\citep{2006ApJ...648L..51S,2009A&A...496..495K}. The dissipative
region (finite-$\sigma$) is determined to be along the
magnetic-field-lines (outside the LC) that originate outside a
certain fraction $1-w$ of the polar-cap rim radius
(Fig.~\ref{fig01}b).

The above treatment ensures that the $E_{\rm acc}$-values are
consistent with the global solution. For low-$\sigma$-values
($\sigma\lesssim 1\Omega$) $E_{\rm acc}$ saturates locally to some
$E_{\rm max}$-value that depends on the assumed gap-width (i.e.
$w$). For high-$\sigma$-values ($\sigma\gtrsim 10\Omega$) where the
current-density $J$ approaches the corresponding FF-value $E_{\rm
acc}\propto\sigma^{-1}$. The integration for high-$\sigma$-values
becomes cumbersome because of the stiff nature of the resistive
term. Thus, for high-$\sigma$-values ($>10\Omega$) we use the
results for $\sigma_0=10\Omega$ and scale $E_{\rm acc}$ according to
\begin{equation}
    \label{eccscale}
    E_{\rm acc}= E_{\rm acc}^{0}\frac{\sigma_0}{\sigma}\;\;\;\;\;\;
    (\sigma>\sigma_0)
\end{equation}
where $E_{\rm acc}^0$ corresponds to $\sigma_0$. We note that
Eq.~\eqref{eccscale} reproduces the correct $E_{\rm acc}$-behavior
for high-$\sigma$ ($E_{\rm acc}\rightarrow 0~{\rm
for}~\sigma\rightarrow\infty$).

We use simulations with $w=0.1$ (unless noted otherwise) that
resolve the stellar radius $r_{\star}$ and the LC-radius $R_{\rm
LC}$ with 15 and 50 grid-points, respectively.

\begin{figure}[!tbh]
\vspace{0.0in}
  \begin{center}
    \includegraphics[width=1.0\linewidth]{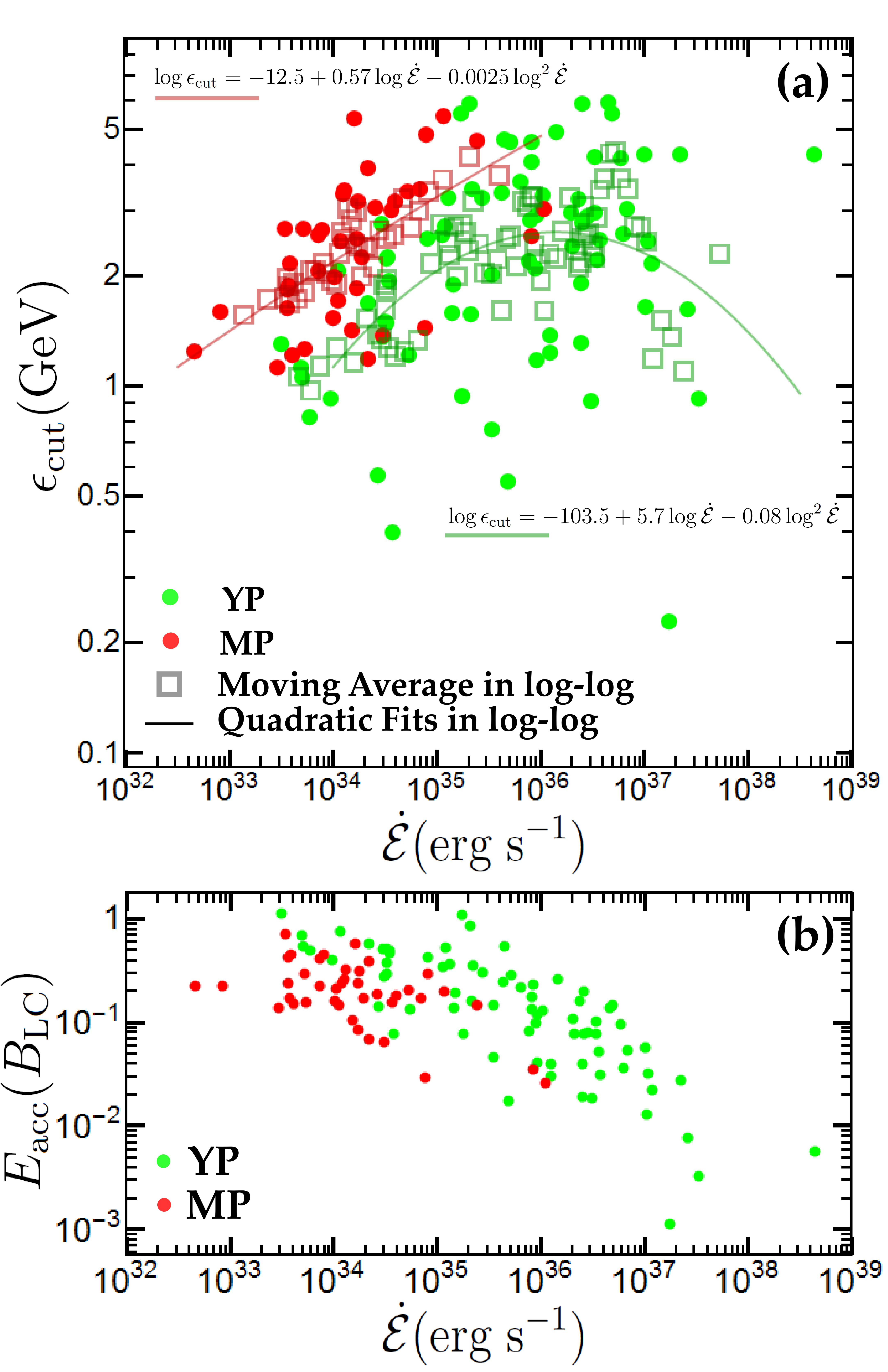}
  \end{center}
  \vspace{0.0in}
  \caption{\textbf{(a)} The \emph{Fermi} $\epsilon_{\rm cut}$ (full-circles) for YPs and MPs.
  The open-squares show the moving-average values (every 4-\emph{Fermi}-objects)
  while the solid-lines show the corresponding quadratic fits.
  \textbf{(b)} The $E_{\rm acc}$ (in $B_{\rm LC}$ units) vs. $\dot{\mathcal{E}}$ for
  \emph{Fermi}-pulsars assuming CR at RRLR near the ECS at the LC.}
  \label{fig02}
  \vspace{-0.1in}
\end{figure}

\section{Guided by \emph{Fermi}: pulsar cutoff-energies}

2PC provides the total $\gamma$-ray luminosities ($L_{\gamma}$) and
the phase-averaged $\epsilon_{\rm cut}$ for most of the \emph{Fermi}
$\gamma$-ray pulsars. However, the $L_{\gamma}$-values depend on the
assumed beaming-factor $F_b$ and the estimated distances. The large
spread in $L_{\gamma}$ with $\dot{\mathcal{E}}$ indicates that other
factors (i.e. $\alpha$-values, variability of $F_b$ with
observer-angle) play an important role on their determination. On
the other hand, the range of $\epsilon_{\rm cut}$ is more limited
($\sim 1-6~\rm GeV$), does not suffer from geometry or distance
uncertainties, and depends weakly only on the adopted fit-model.

In Fig.~\ref{fig02}a full-circles show the $\epsilon_{\rm cut}$ vs.
$\dot{\mathcal{E}}$ for both \emph{Fermi}-Young-Pulsars (YP; green)
and \emph{Fermi}-Millisecond-Pulsars (MP; red). The open-squares
denote the moving averages values and the solid-lines are the
corresponding $\log-\log$ quadratic fits. We see that the
$\epsilon_{\rm cut}$ of YPs increase with $\dot{\mathcal{E}}$ up to
$\sim 10^{36}\rm erg\;s^{-1}$ and then they stabilize or even
decrease. On the other hand, the $\epsilon_{\rm cut}$ of MPs present
a monotonic increase for the observed $\dot{\mathcal{E}}$-values.

The \emph{Fermi} $\epsilon_{\rm cut}$-values provide a unique
insight for the determination of the $E_{\rm acc}$ and through this
for $\sigma$. Assuming that the pulsar emission is due to CR at the
radiation-reaction-limit-regime (RRLR), we get
\begin{subequations}
\begin{mathletters}
\label{rrlr1}
\begin{eqnarray}
    \frac{d\gamma_{\rm L}}{dt} =\frac{q_e \upsilon E_{\rm acc}}{m_e
    c^2}-\frac{2q_e^2\gamma_{\rm L}^4}{3R_{\rm C}^2m_e c}\phantom{\hspace{29pt}}\\
    \dot{\gamma}_{\rm L} =0
    \phantom{\hspace{117pt}}
\end{eqnarray}
\end{mathletters}
\end{subequations} \phantom{\hspace{-11pt}}

\noindent where $m_e, q_e, c, \upsilon, \gamma_{\rm L}, R_C$ are the
electron-mass, electron-charge, speed-of-light, particle-speed along
$E_{\rm acc}$, Lorentz-factor, and radius-of-curvature,
respectively. The first term in Eq.~(\ref{rrlr1}a) describes the
energy-gain due to any $E_{\rm acc}$ the particles encounter while
the second term describes the CR-reaction losses. Assuming also that
all the radiative action is near the ECS close to the LC we have an
estimation for the $R_C\approx R_{\rm LC}$ (see
\citetalias{kalap2014}) and then taking into account the
\emph{Fermi} $\epsilon_{\rm cut}$-values (2PC) and the well-known
expression
\begin{equation}
\label{ecutCR} \epsilon_{\rm cut}=\frac{3}{2}c\hbar\frac{\gamma_{\rm
L}^3}{R_C}
\end{equation}
we can get an estimate of the corresponding $\gamma_{\rm L}$-values.
Applying this estimated value to Eqs.~(\ref{rrlr1}a,b) (for
$\upsilon\approx c$) we get a final estimate of $E_{\rm acc}$. In
Fig.~\ref{fig02}b we plot these $E_{\rm acc}$-values (in the
corresponding $B_{\rm LC}$-units) vs. $\dot{\mathcal{E}}$ for all
the \emph{Fermi}-pulsars. The $E_{\rm acc}$ decreases for
high-$\dot{\mathcal{E}}$ and saturates for low-$\dot{\mathcal{E}}$
around a value that is lower than $B_{\rm LC}$.

The result depicted in Fig.~\ref{fig02}b is important not only
because it is based entirely on \emph{Fermi}-data and
simple/fundamental assumptions but also because it anticipates the
dependence of $\sigma$ on $\dot{\mathcal{E}}$.

\FloatBarrier
\begin{table*}[!htb]
\centering
        \begin{tabular}{ccccccc}
            \hline
            & \multicolumn{3}{c|}{Young Pulsars} & \multicolumn{3}{c}{Millisecond Pulsars}\\
            \cline{2-7}\\[-8pt]
            &  $B_{\star}$ & $P$ & $\dot{\mathcal{E}}_{\rm FF}$ & $B_{\star}$ & $P$ & $\dot{\mathcal{E}}_{\rm FF}$ \\
            &  ($10^{11}$G) & (ms) & ($10^{33}\rm erg\;s^{-1}$) & ($10^{8}$G) & (ms) & ($10^{33}\rm erg\;s^{-1}$) \\ \hline
1 & 3.2 & 398.1 & 0.06 & 1.3 & 5.4 & 0.27\\
2 & 8.7 & 398.1 & 0.44 & 1.6 & 5.4 & 0.44\\
3 & 3.2 & 223.9 & 0.57 & 2.0 & 5.4 & 0.69\\
4 & 16.6 & 398.1 & 1.58 & 1.3 & 4.0 & 0.91\\
5 & 8.7 & 223.9 & 4.36 & 1.6 & 4.0 & 1.44\\
6 & 3.2 & 125.9 & 5.75 & 3.2 & 5.4 & 1.73\\
7 & 43.7 & 398.1 & 10.95 & 2.0 & 4.0 & 2.29\\
8 & 16.6 & 223.9 & 15.82 & 1.3 & 3.0 & 3.02\\
9 & 8.7 & 125.9 & 43.58 & 4.5 & 5.4 & 3.46\\
10 & 3.2 & 70.8 & 57.45 & 1.6 & 3.0 & 4.78\\
11 & 104.7 & 398.1 & 62.99 & 3.2 & 4.0 & 5.75\\
12 & 43.7 & 223.9 & 109.47 & 6.5 & 5.4 & 7.23\\
13 & 16.6 & 125.9 & 158.23 & 2.0 & 3.0 & 7.57\\
14 & 251.2 & 398.1 & 362.49 & 1.3 & 2.2 & 9.98\\
15 & 8.7 & 70.8 & 435.81 & 4.5 & 4.0 & 11.46\\
16 & 3.2 & 39.8 & 574.51 & 1.6 & 2.2 & 15.82\\
17 & 104.7 & 223.9 & 629.93 & 3.2 & 3.0 & 19.02\\
18 & 43.7 & 125.9 & 1.1$\times 10^3$ & 6.5 & 4.0 & 23.95\\
19 & 16.6 & 70.8 & 1.6$\times 10^3$ & 2.0 & 2.2 & 25.08\\
20 & 251.2 & 223.9 & 3.6$\times 10^3$ & 1.3 & 1.6 & 33.06\\
21 & 8.7 & 39.8 & 4.4$\times 10^3$ & 4.5 & 3.0 & 37.96\\
22 & 104.7 & 125.9 & 6.3$\times 10^3$ & 1.6 & 1.6 & 52.40\\
23 & 43.7 & 70.8 & 1.1$\times 10^4$ & 3.2 & 2.2 & 62.99\\
24 & 16.6 & 39.8 & 1.6$\times 10^4$ & 6.5 & 3.0 & 79.30\\
25 & 251.2 & 125.9 & 3.6$\times 10^4$ & 2.0 & 1.6 & 83.04\\
26 & 104.7 & 70.8 & 6.3$\times 10^4$ & 4.5 & 2.2 & 125.69\\
27 & 43.7 & 39.8 & 1.1$\times 10^5$ & 3.2 & 1.6 & 208.59\\
28 & 251.2 & 70.8 & 3.6$\times 10^5$ & 6.5 & 2.2 & 262.60\\
29 & 104.7 & 39.8 & 6.3$\times 10^5$ & 4.5 & 1.6 & 416.19\\
30 & 251.2 & 39.8 & 3.6$\times 10^6$ & 6.5 & 1.6 & 869.55\\

        \end{tabular}
    \caption{The model $(B_{\star},P)$-combinations and the corresponding
    FF $\dot{\mathcal{E}}$-values for the aligned rotator
    ($\alpha=0^{\circ}$, Eq.~\ref{spindown}).}
    \label{tab01}
\end{table*}

\section{Finding $\sigma$}

Using the models, described in
Section~\ref{sec2}\footnote{{Actually, for each model, we use a
steady-state snapshot considering it is static in the corotating
frame.}}, we integrate test particle trajectories assuming the
Goldreich-Julian flux $n_{\rm GJ}c$ from the polar-cap. Following an
approach similar to those we used in \citetalias{kalap2014},
\cite{2015ApJ...804...84B} we define particle trajectories
considering that the velocity is everywhere determined by the so
called Aristotelian Electrodynamics (hereafter AE)
\citep{2012arXiv1205.3367G}
\begin{equation}
\label{ae}
\mathbf{v}=\frac{\mathbf{E}\times\mathbf{B}\pm(B_0\mathbf{B}+E_0\mathbf{E})}{B^2+E_0^2}
\end{equation}
where the two signs correspond to the two different types of charge.
We always choose the charge that is accelerated outwards. {The
quantities $E_0$ and $B_0$ are related to the Lorentz invariants
\citep{2008arXiv0802.1716G,2012ApJ...746...60L}
\begin{equation}
\label{e0b0} E_0 B_0=\mathbf{E}\cdot
\mathbf{B},\;\;E_0^2-B_0^2=E^2-B^2
\end{equation}
and $E_0$ is the electric field in the frame where $\mathbf{E}$ and
$\mathbf{B}$ are parallel and is the actual accelerating electric
component which becomes zero only when $\mathbf{E}\cdot
\mathbf{B}=0$ and $E<B$. Equation~\eqref{ae} describes accurately
the asymptotic behavior of the particle velocities and the
corresponding trajectory determination is very close to the real
ones. Apparently, all the velocities in AE are by definition equal
to $c$ i.e. the asymptotic value. This implies that Eq.~\eqref{ae}
can be used only for the determination of the trajectory shape and
that no information about the particles' dynamics/energetics can be
derived by it. Thus, along each of these trajectories we compute
$\gamma_{\rm L}$ by integrating Eq.~(\ref{rrlr1}a) taking into
account the local $R_{C}$-values that are calculated by the
geometric-shapes of the trajectories defined by Eq.~\eqref{ae}.} The
$\gamma_{\rm L},R_{C}$-values allow the derivation of the
corresponding emission. Collecting all the emitted photons we can
construct sky-maps and compute spectra.

In our study, we have used a series of models for different
combinations of $\alpha$-values, periods $(P)$, stellar-surface
magnetic-fields $(B_{\star})$, and $\sigma$-values. The
corresponding FF spin-down rate reads \citep{2006ApJ...648L..51S}
\begin{equation}
    \label{spindown}
    \dot{\mathcal{E}}=\frac{4\pi^4 r_{\star}^6}{c^3}\frac{B_{\star}^2}{P^4}(1+\sin^2\alpha)
    \vspace{-0.05in}
\end{equation}
where $r_{\star}\simeq 10^6\rm cm$ is the stellar-radius.
Table~\ref{tab01} shows the $(P,B_{\star})$-combinations that
produce the entire range of the observed $\dot{\mathcal{E}}$-values
for YPs and MPs.

Moreover, for each of these 30 $\dot{\mathcal{E}}_{\rm FF}$-values
we have considered 4 conductivities $\sigma=(1,10,10^2,10^3)\Omega$
and 18 $\alpha$-values
$(\alpha=5^{\circ},10^{\circ},15^{\circ},\ldots,90^{\circ};
\text{every } 5^{\circ})$. {For $\sigma=10^2\Omega \text{ and }
10^3\Omega$ we use the simulation for $\sigma=10\Omega$ and scale
the accelerating electric field according to the relation
\eqref{eccscale}.} Thus, in total we have $30\times 4\times 18=2160$
YP-models and 2160 MP-models.

\begin{figure*}[!tbh]
\vspace{0.0in}
  \begin{center}
    \includegraphics[width=1.0\linewidth]{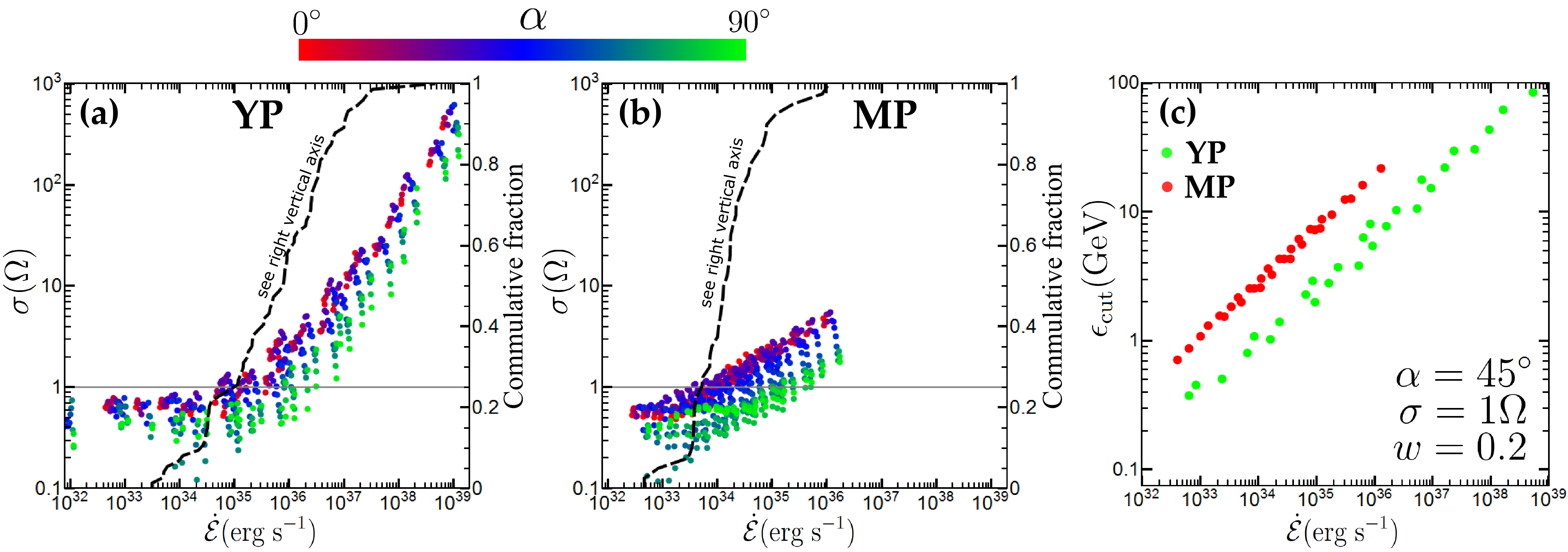}
  \end{center}
  \vspace{0.0in}
  \caption{\textbf{(a)}, \textbf{(b)} The YP and MP model
  $\sigma_{\rm opt}$-values for $w=0.1$
  that reproduce the \emph{Fermi} $\epsilon_{\rm cut}$. The values below the
  gray-lines have been derived by extrapolation (see text for more details)
  and indicate larger $w$-values. The dashed-lines show the cumulative fraction
  of the corresponding \emph{Fermi}-group (right vertical-axes).
  \textbf{(c)} The model $\epsilon_{\rm cut}$-values
  for the indicated ($\alpha,\sigma,w$)-values.}
  \label{fig03}
  \vspace{-0.1in}
\end{figure*}

For each of these models we build the spectrum taking into account
the emission from the entire magnetosphere (up to $r=2.5R_{\rm
LC}$). The resulting spectral-energy-distributions are then fit with
the model used in 2PC, namely,
\begin{equation}
    \label{spfit}
    \frac{dN}{d\epsilon}=A \epsilon^{-\Gamma}\exp\left(-\frac{\epsilon}{\epsilon_{\rm
    cut}}\right)
    \vspace{-0.05in}
\end{equation}
where $\Gamma$ is the photon-index. For each
$(B_{\star},~P,~\alpha)$-combination we compute $\epsilon_{\rm cut}$
for the considered 4 $\sigma$-values. A linear-interpolation of
these $(\log\sigma,\log\epsilon_{\rm cut})$-values is then used to
find the optimum $\sigma_{\rm opt}$-value that reproduces the
$\epsilon_{\rm cut}$ indicated by the fits shown in Fig.\ref{fig02}a
for the corresponding (through Eq.\ref{spindown})
$\dot{\mathcal{E}}$-values.

In Fig.~\ref{fig03}a,b we present these $\sigma_{\rm opt}$-values
for all the YP and MP models, respectively. The $\sigma_{\rm
opt}$-values of the points that are below the $\sigma=1\Omega$
gray-line have been determined by extrapolations of the
linear-interpolations. We note that the dashed-lines indicate the
cumulative fraction of each \emph{Fermi} pulsar-group (YP, MP) that
is observed below the corresponding $\dot{\mathcal{E}}$-value. For
each $\dot{\mathcal{E}}$-value the $\sigma_{\rm opt}$ decreases with
$\alpha$, especially for high $\alpha$-values.

For YPs, $\sigma_{\rm opt}\propto\dot{\mathcal{E}}$ for high
$\dot{\mathcal{E}}$ and saturates towards lower
$\dot{\mathcal{E}}$-values. The $\sigma_{\rm opt}$-values below
$1\Omega$ imply the necessity of higher $E_{\rm acc}$-values than
those found in our models. Nonetheless, these $\sigma$-values are
only slightly lower than $1\Omega$ while they appear close to the
low-end of $\dot{\mathcal{E}}$-values that \emph{Fermi} observes
YPs. In our models, the $E_{\rm acc}$-values do not depend only on
the adopted $\sigma$-value but also on the adopted gap-width (i.e.
$w$). In our modeling, $w$ remains the same for all the
$\dot{\mathcal{E}}$-values. However, smaller $\sigma_{\rm opt}$
indicates that the corresponding model struggles more to eliminate
the $E_{\rm acc}$. This difficulty implies also wider gap-widths
(i.e. higher $w$-values). We tried a few models that have $w=0.2$
(instead of $w=0.1$) and found an increase of $\epsilon_{\rm cut}$
that leads to an increase of the corresponding $\sigma_{\rm
opt}$-value by a factor of $\sim 1.5-2.0$. This small increase is
sufficient to restore most of the points that are below $1\Omega$
(Fig.~\ref{fig03}a) back to $\sigma_{\rm opt}\gtrsim 1\Omega$.

MPs show similar behavior even though they extend over a smaller
$\dot{\mathcal{E}}$-range of rather low $\dot{\mathcal{E}}$-values
(Fig.~\ref{fig03}b). The rising part is less steep ($\sigma\propto
\dot{\mathcal{E}}^{1/3}$) than that of YPs for the high
$\dot{\mathcal{E}}$-values. As mentioned in the previous section,
the $\epsilon_{\rm cut}$ of YPs stop increasing for
$\dot{\mathcal{E}}\gtrsim 10^{36}\rm erg\;s^{-1}$ while the
$\epsilon_{\rm cut}$ of MPs seem to increase for all the observed
$\dot{\mathcal{E}}$-values. Thus, a faster increase of $\sigma_{\rm
opt}$ of YPs is required to reduce the corresponding $E_{\rm acc}$
more efficiently.

Similarly to YPs, a wider-gap is implied for the low
$\dot{\mathcal{E}}$-values of MPs. Wider-gaps mean larger
emission-domains in the magnetosphere which is totally consistent
with the observations (for many \emph{Fermi}-MPs and the
low-$\dot{\mathcal{E}}$ \emph{Fermi}-YPs) that show wider
$\gamma$-ray pulses and, in general, more complex $\gamma$-ray
light-curves (Renault~et~al.~2017, in prep).

\begin{figure}[!tbh]
\vspace{0.0in}
  \begin{center}
    \includegraphics[width=1.0\linewidth]{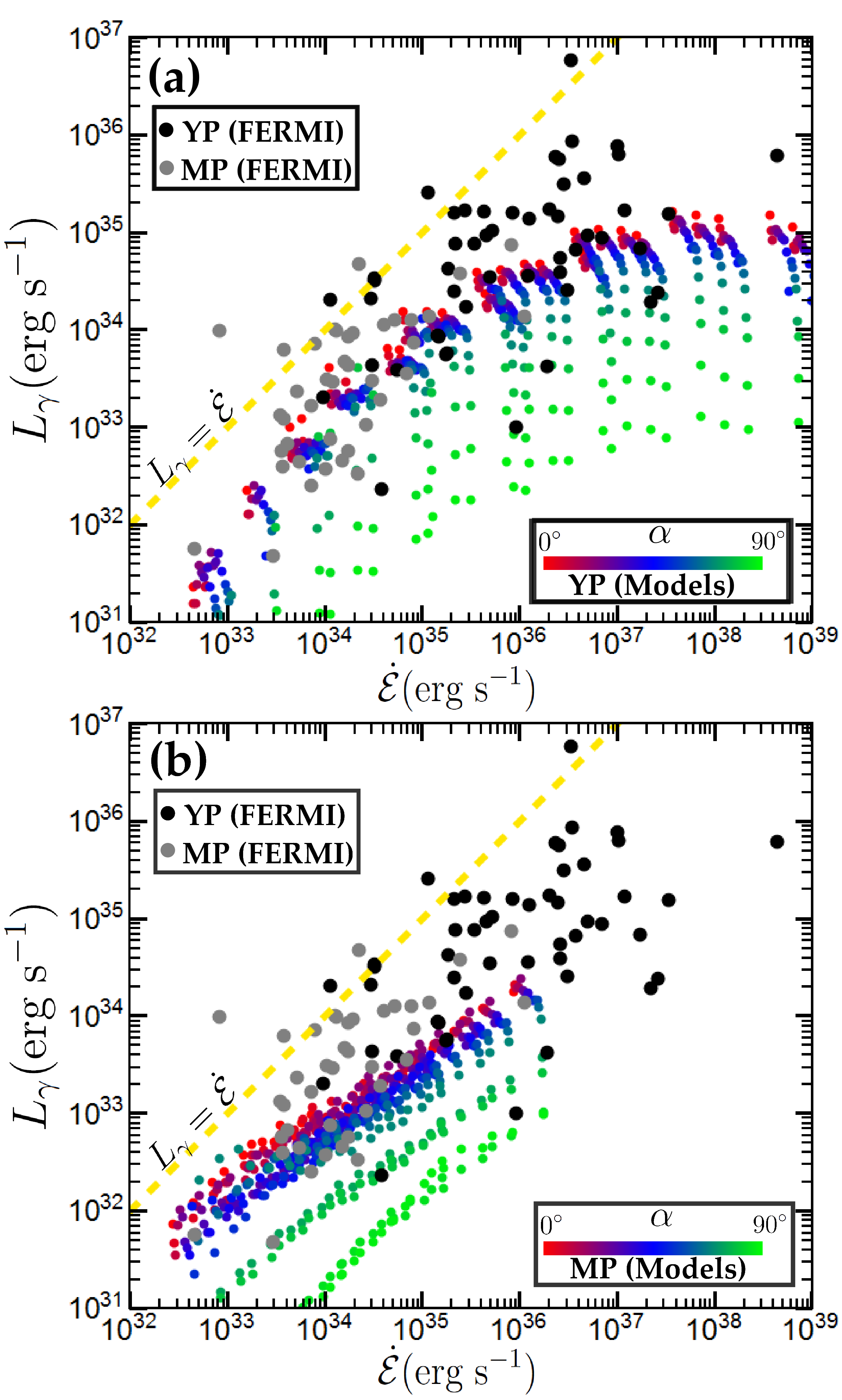}
  \end{center}
  \vspace{0.0in}
  \caption{The model $L_{\gamma}$-values
  (color points) together with the corresponding
  \emph{Fermi}-values as indicated in the Figure.
  The yellow dashed-lines denote 100\% efficiency.
  The comparison between models and observations indicates higher multiplicities
  for the emitting particles at high $\dot{\mathcal{E}}$.}
  \label{fig04}
  \vspace{-0.1in}
\end{figure}

Our analysis provides also a possible explanation for why YPs and
MPs are not observed for $\dot{\mathcal{E}}\lesssim 10^{34}\rm
erg\;s^{-1}$ and $\dot{\mathcal{E}}\lesssim 10^{33}\rm erg\;s^{-1}$,
respectively. We have already seen in Fig.~\ref{fig02}a that the
$\epsilon_{\rm cut}$ of \emph{Fermi} YPs and MPs decrease towards
low $\dot{\mathcal{E}}$-values. In Fig.~\ref{fig03}c we plot the
$\epsilon_{\rm cut}$ vs. $\dot{\mathcal{E}}$ for all the models for
$\alpha=45^{\circ}$, $\sigma=1\Omega$, and $w=0.2$. Below some
$\dot{\mathcal{E}}$ the corresponding $\epsilon_{\rm cut}$ becomes
small, approaching the \emph{Fermi}-threshold ($\sim0.1\rm GeV$),
which, in combination of lower luminosities, apparently makes their
detection more difficult.

In Appendix~\ref{append} \citep[see also][]{2013arXiv1309.6974G}, we
show that assuming emission due to CR at RRLR (for the same $\sigma,
w$) we get
\begin{equation}
    \label{ecutedot}
    \epsilon_{\rm cut}\propto\dot{\mathcal{E}}^{3/8}.
\end{equation}
The slope 3/8 is followed very well by the model data-points in
Fig.~\ref{fig03}c. We show also that the $\epsilon_{\rm cut}$-values
for pulsars of the same $\dot{\mathcal{E}}$, but of different
$(B_{\star},P)$-combinations, read
\begin{equation}
    \label{ecutbstar}
    \epsilon_{\rm cut}\propto B_{\star}^{-1/8}.
\end{equation}
Applying the previous expression to \emph{Fermi} MPs and YPs, taking
into account that ${B_{\star_{\rm MP}}}\approx 10^{-4}{B_{\star_{\rm
YP}}}$ we get that ${\epsilon_{\rm cut_{\rm
MP}}}\approx{3\epsilon_{\rm cut_{\rm YP}}}$ which is followed
exactly by the models (Fig.~\ref{fig03}c). This rule explains also
why \emph{Fermi}-MPs have, on average, higher $\epsilon_{\rm
cut}$-values than YPs for the same $\dot{\mathcal{E}}$. The actual
average $\epsilon_{\rm cut}$-ratios, as can be derived by the data
shown in Fig.~\ref{fig02}a, vary by a factor $\sim 1.5-2$ $(<3)$
mainly because of the slightly different values of $\sigma$ and $w$.

Finally, for completeness, in Fig.~\ref{fig04}a,b we present the
model $L_{\gamma}$ vs. $\dot{\mathcal{E}}$ corresponding to the
$\sigma_{\rm opt}$-values ($w=0.1$) for YPs and MPs, respectively.
We note that for the cases that $\sigma_{\rm opt}<1\Omega$
(Fig.~\ref{fig03}a,b) we plot the extrapolated $L_{\gamma}$-values
of the linear-interpolation of $(\log\sigma,\log L_{\gamma})$ we
have for $\sigma>1\Omega$. Figure~\ref{fig04} shows that
$L_{\gamma}$ decreases with $\alpha$ with the dependence on $\alpha$
becoming stronger at high $\alpha$-values mainly because of the
lower particle-fluxes $n_{\rm
GJ}\propto\mathbf{B}\cdot\mathbf{\Omega}\propto \cos(\alpha)$ which
implies higher relative multiplicities for higher $\alpha$.

In the previous section, we discussed the uncertainties of the
observed \emph{Fermi} $L_{\gamma}$-values. From the model
point-of-view the main uncertainty is the number of particles that
accelerate at every point of the magnetosphere and contribute to the
high-energy emission. Assuming a GJ-flux for all the models we see
that the intermediate and low-$\alpha$ of both YPs and MPs are able
to reproduce the observed $L_{\gamma}$-values at
low-$\dot{\mathcal{E}}$ even though they never reach close to the
100\% efficiency (yellow dashed-line). For higher
$\dot{\mathcal{E}}$, the model $L_{\gamma}$-values are lower than
the observed ones with the effect being more prominent for the YPs.
The \emph{Fermi} $L_{\gamma}$-values of YPs increase slower with
$\dot{\mathcal{E}}$, at high $\dot{\mathcal{E}}$, implying lower
$\gamma$-ray efficiency. The YP-models show a similar behavior
although their $L_{\gamma}$-values increase much slower and seem
even to decrease at very high $\dot{\mathcal{E}}$. This discrepancy
can be reconciled by assuming an increased particle multiplicity for
$\dot{\mathcal{E}}\gtrsim~10^{36}\rm erg\;s^{-1}$; such an
assumption is in agreement with the increase in $\sigma$ (which is
supposedly attributed to higher particle-multiplicities) above this
$\dot{\mathcal{E}}$-value shown in Fig.\ref{fig03}a. We note also
that the $L_{\gamma}$-inconsistency at the relatively higher
$\dot{\mathcal{E}}$-values for MPs (Fig.~\ref{fig04}b) appears
milder because the corresponding $\sigma_{\rm opt}$ increase is
smaller (Fig.~\ref{fig03}b).

\section{Conclusions}

In this paper, by expanding our previous studies, we interpret the
Fermi pulsar $\gamma$-ray phenomenology within the framework of
sophisticated dissipative pulsar magnetosphere models.

We refine our FIDO models by restricting the dissipative regions to
near the ECS. For this, we run simulations that have
magnetic-field-line dependent conductivity. This approach allows the
exploration of low $\sigma$-values providing $E_{\rm acc}$ that are
consistent with the global structure. Although these solutions are
dissipative, the corresponding field-structure remains close to the
FF one which is necessary for the reproduction of the nice
$\delta-\Delta$ correlation (2PC; \citetalias{kalap2014}).

Moreover, based on very basic assumptions that the observed
$\gamma$-ray emission
\begin{enumerate}[(a)]
    \item is due to CR at RRLR
    \item is produced at the ECS,
    near the LC,
\end{enumerate}
we show that the \emph{Fermi} $\epsilon_{\rm cut}$-values reveal the
required $E_{\rm acc}$-values (in $B_{\rm LC}$ units) which decrease
with $\dot{\mathcal{E}}$, at high $\dot{\mathcal{E}}$, while they
stabilize at low $\dot{\mathcal{E}}$, below $B_{\rm LC}$.

Motivated by the previous result and taking into account the
\emph{Fermi} $\epsilon_{\rm cut}$-variation with $\dot{\mathcal{E}}$
we derive, for two series of models that cover the entire range of
the observed $\dot{\mathcal{E}}$ of YPs and MPs, the different
$\sigma_{\rm opt}$-values that reproduce the corresponding
$\epsilon_{\rm cut}$. We find that the $\sigma_{\rm opt}$ increase
with $\dot{\mathcal{E}}$, at high $\dot{\mathcal{E}}$. For the low
$\dot{\mathcal{E}}$ the models struggle to produce the observed
$\epsilon_{\rm cut}$-values indicating the need for larger
dissipative regions that can provide the slightly higher $E_{\rm
acc}$ needed in these cases.

The comparison between the model-$L_{\gamma}$ with the observed
values becomes difficult because of the existing uncertainties in
both the 2PC data (i.e. pulsar distances, unknown beaming-factors)
and the models (i.e. multiplicity of the emitting particles).
However, comparing mainly the trends of the $L_{\gamma}$-dependence
on $\dot{\mathcal{E}}$, it becomes clear that relatively higher
emitting particles multiplicities are needed for high
$\dot{\mathcal{E}}$-models and the very high $\alpha$-values.

We emphasize that the seemingly unbiased initial choice of the
model-parameters (i.e. same size of the dissipative region, same
emitting particle-multiplicity) independent of the
$\dot{\mathcal{E}}$, led to some problems, the solutions of which
are consistent with the underlying theoretical view. Thus, the
emerging necessities of larger dissipative regions towards
low-$\dot{\mathcal{E}}$ and of higher emitting
particle-multiplicities towards high-$\dot{\mathcal{E}}$ are
consistent with the lower (higher) $\sigma_{\rm opt}$ at low (high)
$\dot{\mathcal{E}}$ and the associated lower (higher) pair
production efficiency, respectively. {We note also that even though
our simple consideration of only two regimes of conductivity (finite
$\sigma$ near the ECS and infinite everywhere else) is successful in
interpreting the observations, in reality the situation is expected
to be more complex. Thus, a possible generalization might be a
gradual variation of the conductivity with the polar cap radius
(i.e. polar-angle from the magnetic pole) and the spherical radius.}

Our models guided by observations provide a complete macroscopic
picture with meaningful constraints that deepens our understanding
about the pulsar $\gamma$-ray emission mechanisms and shows that CR
can provide the observed \emph{Fermi} pulsar-emission, in contrast
to models that advocate synchrotron-emission at GeV energies
\citep{2012MNRAS.424.2023P,2016MNRAS.457.2401C}. However, they are
not self-consistent in the sense that they cannot provide
unambiguous information about the microscopic properties of the
magnetospheric plasma, such as pair-creation and particle
distribution function. This kind of studies require the use of
kinetic particle-in-cell simulations
\citep{2014ApJ...785L..33P,2014ApJ...795L..22C,2015NewA...36...37B,2015ApJ...801L..19P,2016MNRAS.457.2401C}
and are expected to reveal the dependence of the macroscopic
parameters found in the present study on the microphysical processes
of pulsar magnetospheres. We have started exploring this research
path and we will present our results in forthcoming papers.

\acknowledgments

We would like to thank an anonymous referee for helpful suggestions
that improved the paper. This work is supported by the National
Science Foundation under Grant No. AST-1616632 and by the NASA
Astrophysics Theory Program. Resources supporting this work were
provided by the NASA High-End Computing (HEC) Program through the
NASA Advanced Supercomputing (NAS) Facility at NASA Ames Research
Center and NASA Center for Climate Simulation (NCCS) at NASA Goddard
Space Flight Center.

\appendix
\section{A.}
\label{append}

We assume emission at the LC near the ECS due to CR at RRLR for
models of specific $w$ and $\sigma$. Then $E_{\rm acc}\propto B_{\rm
LC}\propto B_{\star}R_{\rm LC}^{-3}$ and because $R_{\rm LC}\propto
P$
\begin{equation}\label{app01}
    E_{\rm acc}\propto B_{\star}P^{-3}
\end{equation}
and from $\dot{\mathcal{E}}\propto B_{\star}^2P^{-4}$
(Eq.~\ref{spindown})
\begin{equation}\label{app02}
    E_{\rm acc}\propto \dot{\mathcal{E}}^{1/2}P^{-1}.
\end{equation}
Equation~\eqref{rrlr1} gives $\gamma_{\rm L}\propto E_{\rm
acc}^{1/4} R_{\rm C}^{1/2}$ and because $R_{\rm C}\propto R_{\rm
LC}\propto P$
\begin{equation}\label{app03}
    \gamma_{\rm L}\propto E_{\rm acc}^{1/4}P^{1/2}
\end{equation}
and using Eq.~\eqref{app02}
\begin{equation}\label{app04}
    \gamma_{\rm L}\propto \dot{\mathcal{E}}^{1/8}P^{1/4}.
\end{equation}
From Eq.~\eqref{ecutCR} we have also
\begin{equation}\label{app05}
    \epsilon_{\rm cut}\propto \gamma_{\rm L}^3P^{-1}
\end{equation}
and using Eq.~\eqref{app04}
\begin{equation}\label{app06}
    \epsilon_{\rm cut}\propto \dot{\mathcal{E}}^{3/8}P^{-1/4}.
\end{equation}
Taking into account that for each pulsar group (YP, MP) the range of
the observed $\dot{\mathcal{E}}$-values is much broader than that of
$P$-values it becomes clear that for pulsars $\epsilon_{\rm
cut}\propto \dot{\mathcal{E}}^{3/8}$ while the weak dependence on
$P$ produces just a small spread of the $\epsilon_{\rm cut}$-values.

Moreover, for pulsars of the same $\dot{\mathcal{E}}$, $P\propto
B_{\star}^{1/2}$ (Eq.~\ref{spindown}) and thus, from
Eqs.~\eqref{app01},\eqref{app03}, and \eqref{app05}
\begin{equation}\label{app07}
    \epsilon_{\rm cut}\propto B_{\star}^{-1/8}.
\end{equation}
%




\end{document}